\documentclass[12pt,tightenlines,eqsecnum,floats,aps,amsmath,amssymb,nofootinbib,prd,floatfix,superscriptaddress]{revtex4}

\usepackage{setspace}
\usepackage{subfigure}
\usepackage{amsmath,amssymb,amsfonts,amsthm,mathrsfs}
\usepackage{graphicx,wrapfig}
\usepackage{enumerate} 

\def\f{\frac}
\def\l{\left}
\def\r{\right}

\def\d{\textrm{d}}

\newcommand{\be}{\nopagebreak[3]\begin{equation}}
\newcommand{\ee}{\end{equation}}
\newcommand{\bea}{\nopagebreak[3]\begin{eqnarray}}
\newcommand{\ea}{\end{eqnarray}}

\begin{document}

\title{Inflation from inhomogeneous polarized Gowdy model}

\author{Javier Olmedo}
 \email{javolmedo@ugr.es}
\affiliation{Departamento de F\'isica Te\'orica y del Cosmos, Universidad de Granada, Granada-18071, Spain}%

\begin{abstract}
We study polarized Gowdy cosmologies on the three torus coupled to a massive scalar field. The phase space of the model admits a simple splitting between homogeneous and inhomogeneous sectors after a suitable gauge fixing. The presence of the mass term of the scalar field breaks the linearity of the equations of motion of the inhomogeneous fields. We discuss regimes of physical interest in which we recover a linear dynamics of these nonperturbative inhomogeneities, despite the metric is fully inhomogeneous at early times. We expand the inhomogeneous fields in Fourier modes and express them at all times as linear combinations of a basis of orthonormal complex solutions to the equations of motion, with coefficients that turn out to be an infinite collection of constants of motion. We argue that the resulting model can describe a nonperturbative inhomogeneous early universe dominated by the kinetic energy of an inflaton at early times that can eventually reach a slow-roll regime with a nearly exponential expansion at late times that isotropices and homogenizes the geometry.
\end{abstract}
\maketitle
\section{Introduction}

The standard cosmological model successfully explains in a reasonable simple way most of the features of the observable universe, from the hot Big Bang nucleosynthesis until today. It is actually based on the cosmological principle, namely, that our Universe began being homogeneous and isotropic. The origin of small seeds at early times precluding the inhomogeneities that we observe today is explained within the inflationary paradigm. Here, an early exponential expansion of the universe can squeeze quantum vacuum fluctuations and stretch them from microscopical to large macroscopical scales. However, one can ask the question of whether the inflationary paradigm requires the cosmological principle as a fundamental ingredient, or if it is actually a natural consequence of an early exponential expansion. For instance, if spacetime anisotropies are present at early times, a geometry expanding sufficiently fast can dilute them (whenever they are not sourced by some anisotropic stress-energy tensor). However, despite cosmological expansion dilutes spacetime anisotropies, they still leave imprints in the power spectrum of primordial perturbations \cite{Pereira:2007,Pitrou:2008}, as it was shown in detail in a concrete inflationary scenario \cite{Agullo:2020-lett,Agullo:2020-class,Agullo:2020-lqc}. On the other hand, inhomogeneous spacetimes coupled to matter have been studied in order to discern if inhomogeneities can prevent inflation to occur \cite{Goldwirth:1991}. For instance, spherically symmetric inhomogeneous spacetimes will not enter the inflationary regime if inhomogeneities are large enough at the onset of inflation \cite{Goldwirth:1989,Calzetta}. On the other hand, spacetimes with inhomogeneities in one direction \cite{kurki1987inflation} do not seem to prevent inflation. More recent analyses \cite{East:2015,Clough:2016} within full numerical relativity and more general choices of initial inhomogeneities show that an exponential expansion can occur even if gradients in the inflaton field dominate over its potential energy. 

Although inhomogeneous scenarios have been a matter of considerable research, there is a popular family of inhomogeneous cosmological spacetimes that remain quite unexplored in the context of inflation: the so-called Gowdy cosmologies \cite{gowdy1971gravitational,gowdy1974vacuum}. These correspond to spacetimes with compact spatial sections and two commuting spacelike Killing vector fields. In vacuum, these spacetimes display inhomogeneities in one spatial direction that correspond to the two possible polarizations of propagating (nonlinear) gravitational waves. In the case in which the two Killing vector fields are orthogonal, one is left with only one of the polarizations \cite{moncrief1981global,moncrief1981infinite,isenberg1990asymptotic}. These are the so-called polarized Gowdy models (also denoted as Einstein-Rosen-Gowdy metrics when the spatial topology is flat). It is interesting to note that there is a choice of field variables and a gauge fixing (i.e. a choice of spacetime slicing) where the dynamics of the inhomogeneous field is described by a second order, linear Klein-Gordon equation, despite the spacetime metric is nonlinear (and nonlocal) in this field. The simplicity of this setting motivated several studies on its quantization \cite{misner1973,berger1974,berger1975,berger1984,marugan1997,MenaMarugan:1998,corichi2002,pierri2002,torre2002,cortez2005,corichi2006,corichi2006b,corichi2006c,corichi2007,cortez2007,MartinBenito:2008ej,Garay:2010sk,MartinBenito:2010bh,deBlas:2017goa}. Moreover, the inhomogeneous degrees of freedom still satisfy linear equations of motion even if one couples the system to an inhomogeneous massless scalar field with the same symmetries of the spacetime \cite{BarberoG:2007qra,BarberoG.:2007rd,martin2011,Martin-Benito:2013jqa,Navascues:2014rla,Navascues:2015saa}. 

Since most of the research on these spacetimes have been focused on the study of the quantization of the system, with few observational applications, in this work we propose to fill this gap by adding a mass term for the scalar field. This setting, in the homogeneous limit, admits inflationary solutions of interest in current observational cosmology. However, if large inhomogeneities are present, it is unclear if the spacetime will develop an exponential expansion at late times. Besides, the addition of a potential will break the linearity of the equations of motion of the inhomogeneous degrees of freedom. We will discuss the situations in which a linear dynamics is approximately recovered, even when inhomogeneities are not negligible. Within this linear approximation, we characterize the space of solutions and the physical degrees of freedom of the inhomogeneities. This analysis opens the possibility of exploring the observational implications of a classical and quantum treatment of the system, including the inhomogeneous degrees of freedom, within the inflationary paradigm.

The paper is organized as follows. In Sec. \ref{sec:red-model} we introduce the reduced model after a partial gauge fixing. Then we discuss the linear regime of the model in Sec. \ref{sec:linear}. In particular, we propose a complete gauge fixing and derive the equations of motion of the system with a characterization of the solutions space of the inhomogeneous degrees of freedom that additionally allows us to identify the corresponding set of Dirac observables that parametrize the initial state of this inhomogeneous sector. We draw up our conclusions in Sec. \ref{sec:concl}. For the sake of completeness, we added four appendixes where review the symmetry reduction (Appendix \ref{app:sym-gauge-red}), the gauge fixing of the model (Appendix \ref{sec:gauge-fix}), discuss its isotropic sector (Appendix \ref{app:iso-sys}) and provide explicitly its equations of motion  (Appendix \ref{app:eom}).

\section{Symmetry reduced classical system}\label{sec:red-model}

We will start with the Einstein-Hilbert action minimally coupled to a scalar field $\Phi$ that evolves under the influence of a potential $V(\Phi)$ (quadratic in the field). The spacetime manifold will be $M=\mathbb{R} \times T^3$, with $T^3$ the flat three torus. We will adopt an Arnowitt-Deser-Misner (ADM) formulation \cite{arnowitt1962dynamics}. Here, the phase space 
of general relativity  is characterized by two  couples of fields defined on  $T^3$,    $(h_{ij}(\vec{x}),\pi^{ij}(\vec{x});\Phi(\vec{x}),P_{\Phi}(\vec{x}))$, where $h_{ij}(\vec{x})$ is a Riemannian metric that describes the intrinsic spatial geometry of $T^3$, with its conjugate momentum  $\pi^{ij}(\vec{x})$ describing  its extrinsic geometry (Latin indices $i,j$ run from 1 to 3), and $P_{\Phi}(\vec{x})$ is the conjugate momentum of the scalar field $\Phi(\vec{x})$. The non-vanishing Poisson brackets between these fields are
\be \label{PB} \{ \Phi(\vec{x}),P_{\Phi}(\vec{x}')\}=\delta^{(3)}(\vec{x}-\vec{x}')\, ,\hspace{1cm}
\{ h_{ij}(\vec{x}),\pi^{k l}(\vec{x}')\}=\delta_{(i}^k\delta_{j)}^{l}\delta^{(3)}(\vec{x}-\vec{x}')\, .\ee
where $\delta_{(i}^k\delta_{j)}^{l}\equiv \frac{1}{2} (\delta_{i}^k\delta_{j}^{l}+\delta_{j}^k\delta_{i}^{l})$.  These canonical fields are subject to the four constraints of general relativity: The scalar and diffeomorphism (or vector) constraints
\bea \label{scons} \mathbb S (\vec{x})&=&\f{2\kappa}{\sqrt{h}}\l( \pi^{ij}\pi_{ij} -\f{1}{2}\pi^2 \r) - \f{\sqrt{h}}{2\kappa} ~^{(3)}R + \f{1}{2\sqrt{h}} P_{\Phi}^2 + \sqrt{h} \, V(\Phi) + \f{\sqrt{h}}{2} D_i \Phi D^i \Phi \approx 0 ,  \\ \label{vcons}
\mathbb V_i(\vec{x})&=&  -2 \sqrt{h}\,  h_{ij} \, D_k(h^{-1/2} \pi^{kj}) +P_{\Phi} \, D_i \Phi \approx 0 , \ea
where the symbol $\approx$ means that they are constrained to vanish on solutions. Here, $\kappa=8\pi G$, $h$ is the determinant of the spatial metric, $^{(3)}R$ the Ricci scalar of the three dimensional hypersurfaces, and $D_i$ the covariant derivative associated with the metric $h_{ij}$.

Time evolution 
is generated by the Hamiltonian 
\be \label{ham} \mathcal H=\int_{T^3} \d^3x \, \Big[N(\vec{x}) \,  \mathbb S (\vec{x})+N^i(\vec{x}) \, \mathbb V_i (\vec{x})\Big]\, ,\ee
which  is a combination of constraints. Here, $N(\vec{x})$ and $N^i(\vec{x})$ are the so-called lapse and shift functions, respectively, and they play the role of Lagrange multipliers. In total, the spacetime metric takes the form
\be\label{eq:gmunu}
ds^2=-(N^2-h_{ij}N^iN^j)dt^2+2N_idx^idt+h_{ij}dx^idx^j.
\ee

After the symmetry reduction and partial gauge fixing (see Appendix \ref{app:sym-gauge-red}), the reduced spacetime metric of polarized Gowdy spacetimes takes the form
\bea\label{eq:red-ds2}\nonumber
&&ds^2=\left\{-[(a_2a_3)^2 \underline N_0^2-(N_0^x+N_1^x)^2]dt^2+2[N^x_0+N^x_1]dxdt+dx^2\right\}\frac{\tilde h}{(a_2a_3)^2}\\
&&+a_2^2e^{-\frac{\xi_1}{\sqrt{a_2a_3}}}dy^2+a_3^2e^{\frac{\xi_1}{\sqrt{a_2a_3}}}dz^2.
\ea
where $t\in \mathbb{R}$, $x\in[0,L_x)$, $y\in[0,L_y)$ and $z\in[0,L_z)$ are time and spatial coordinates, respectively.\footnote{In the usual treatments the coordinates $(x,y,z)$ are denoted as  $(\theta,\delta,\sigma)$.} Besides, $a_I$ with $I=1,2,3$ are three homogeneous phase space variables that, in absence of inhomogeneities, correspond to the three directional scale factors of Bianchi I cosmologies.\footnote{The definition of the homogeneous and inhomogeneous sectors is mathematically motivated by the symplectic structure of the reduced theory and yields a simple splitting of the phase space into homogeneous and inhomogeneous sectors. Here, the directional scale factors $a_I$ are defined in Eq. \eqref{eq:hom-to-ai} out of the homogeneous contributions $\gamma_0$, $\tau_0$ and $\xi_0$ of the fields parametrizing the spatial metric components given in Eq. \eqref{eq:gi}, namely, $\gamma_0 = \frac{1}{L_x}\int_0^{L_x}dx\,\gamma(x)$, \;$\tau_0 = \frac{1}{L_x}\int_0^{L_x}dx\,\tau(x)$ \; and \; $\xi_0 = \frac{1}{L_x}\int_0^{L_x}dx\,\xi(x)$. This allows us to express, for instance, $\gamma(x)=\gamma_0+\gamma_1(x)$, with the latter contribution being the inhomogeneous counterpart of $\gamma(x)$ ---and similarly for the fields $\tau(x)$ and $\xi(x)$, their conjugate momenta, and the matter sector.} Moreover, $\underline N_0$ is the homogeneous contribution of the densitized lapse function, $\xi_1(x)$ represents the (inhomogeneous) gravitational wave content of the geometry, 
\be\label{eq:deth}
\tilde h(x)=(a_1a_2a_3)^2e^{-\frac{\xi_1^2(x)}{4a_2a_3}+\frac{2}{a_1p_{a_1}}\left[\frac{a_2p_{a_2}-a_2p_{a_2}}{2\sqrt{a_2a_3}}\xi_1(x)-P_{\phi_0}\phi_1(x) -\int_0^x \,d\tilde x[\pi_{\xi_1}(\tilde x) \xi_1'(\tilde x)+P_{\phi_1}(\tilde x)\phi_1'(\tilde x)]\right]},
\ee
is the determinant of the spatial metric, and $N_0^x$ and $N_1^x(x)$ are the homogeneous and inhomogeneous counterparts of the shift vector, respectively, with
\be\label{eq:shift-red}
[\underline N_1^x(x)]' = \frac{\underline N_0 }{2a_1p_{a_1}}\tilde h(x) M^2[\phi_0+\phi_1(x)]^2-\frac{1}{L_x}\int_0^{L_x}d\tilde x\frac{\underline N_0}{2a_1p_{a_1}}\tilde h(\tilde x) M^2[\phi_0+\phi_1(\tilde x)]^2.
\ee
Besides, the prime in phase space variables means partial derivation with respect to $x$. In addition, $\phi_0$ and $\phi_1(x)$ represent the homogeneous and inhomogeneous contributions to the scalar field (inflaton) of mass $M$, respectively, and the remaining quantities are the conjugate momenta of each configuration coordinate. Concretely, each pair of conjugate homogeneous and inhomogeneous fields are coordinates in phase space, with Poisson brackets  
\bea \label{PB-red}
\{a_I,p_{a_J}\}&=&\frac{1}{L_xL_yL_z}\delta_{IJ}\, , \quad  \{ \phi_0,P_{\phi_0}\}=\frac{1}{L_xL_yL_z}\, ,
\\ \nonumber \{ \xi_1(x),\pi_{\xi_1}(x')\}&=&\frac{1}{L_yL_z}\delta(x-x')\, ,\quad \{ \phi_1(x),\pi_{\phi_1}(x')\}=\frac{1}{L_yL_z}\delta(x-x')\,,
\ea
where $L_x$, $L_y$ and $L_z$ are coordinate lengths of spatial coordinates $x$, $y$ and $z$ well adapted to the symmetries of the spatial metric.\footnote{Although $L_x$, $L_y$ and $L_z$ are arbitrary coordinate lengths, each choice gives a different physical spacetime.}

The dynamics is governed by the reduced total Hamiltonian  
\be\label{eq:red-totham}
\underline{\cal H}^{(2)} = \underline N_0^x\underline V_x^{(2)}+\underline N_0\underline S^{(2)}.
\ee
with
\bea\nonumber
&&\underline S_0^{(2)} =L_xL_yL_z \bigg[\frac{\kappa}{4}\bigg(a_1^2p_{a_1}^2+a_2^2p_{a_2}^2+a_3^2p_{a_3}^2-2a_1p_{a_1}a_2p_{a_2}-2a_1p_{a_1}a_3p_{a_3}-2a_2p_{a_2}a_3p_{a_3}\bigg)+\frac{P^2_{\phi_0}}{2}\bigg]\\\label{eq:red-hamconst}
&&+L_yL_z \int_0^{L_x}dx\bigg[\kappa \bigg(\frac{a_1^2p_{a_1}^2}{16a_2a_3}\xi_1^2+a_2a_3\pi_{\xi_1}^2\bigg)+\frac{a_2a_3}{4\kappa}(\xi_1')^2+\frac{P^2_{\phi_1}}{2}+\frac{(a_2a_3)^2}{2}(\phi_1')^2+\tilde h V(\phi)\bigg]
\ea 
the reduced scalar constraint, and
\bea\label{eq:red-diff}
\underline V_x^{(2)} &=&L_yL_z \int_{0}^{L_x}\,dx\bigg[\pi_{\xi_1} \xi_1'+P_{\phi_1}\phi_1'\bigg],
\ea
the reduced diffeomorphism constraint. Note that there are no inhomogeneous diffeomorphism and scalar constraints in this reduced theory since they have been removed via gauge fixing (see Appendix \ref{app:sym-gauge-red}).

We should note that, on the one hand, if inhomogeneities vanish, and after setting the shift vector to zero, the resulting spacetime is a Bianchi I model coupled to a massive scalar field.\footnote{In this situation $\int_0^{L_x}dx\tilde h V(\phi)=L_xM^2\phi_0^2/2$.} On the other hand, when inhomogeneities are present and the scalar field is massless (situations that have been analyzed in the literature \cite{BarberoG:2007qra,BarberoG.:2007rd,martin2011,Martin-Benito:2013jqa}), the scalar and diffeomorphism constraints are quadratic in the fields that represent the inhomogeneities. Hence, the equations of motion of the inhomogeneous sector turn out to be linear, despite the spacetime line element is not a linear (not even a local) function of these inhomogeneous fields. Nevertheless, in the massive case, the inhomogeneous sector is not linear anymore. The mass term, which is quadratic in $\phi_1(x)$, is minimally coupled via the determinant of the metric $\tilde h(x)$, which is not homogeneous.\footnote{It is worth to mention that other potentials which are not quadratic in the scalar field will involve additional nonlinear terms in $\phi_1(x)$. However, if the homogeneous contribution of the scalar field dominates over its inhomogeneous part, and if the potential has a minimum, one can still expand the potential in powers of the inhomogeneities up to quadratic order, and recover a linear regime for inhomogeneities. However, this treatment requires extra care about the stability of this sector upon evolution. Therefore, we will leave it for future investigations.} However, in the next section, we will discuss physically interesting situations in which this contribution is either negligible or nearly homogeneous. In these regimes we recover an approximate linear behavior in the equations of motion of the inhomogeneities.  

\section{Linear inhomogeneous sector: approximations, Fourier expansion and classical dynamics}\label{sec:linear}

As we mentioned, the reduced scalar constraint is (approximately) quadratic in the inhomogeneities in at least two interesting situations: i) the potential of the scalar field $V(\phi)$ is negligible, or ii) the determinant of the spatial metric $\tilde h(x)$  is (approximately) homogeneous. From the observational point of view, this potential is disfavoured by present constraints in the tensor-to-scalar ratio \cite{planck2021}. However, from the conceptual perspective and the purpose of this research, a quadratic potential is the most convenient choice. The mass of the inflaton has been constrained by observations to be around $M\simeq 1.2\cdot 10^{-6}$ (in Planck units) in the homogeneous limit. We will assume this value is still valid in presence of inhomogeneities. Therefore, if we start in a nearly Planckian spacetime that is not dominated by the potential of the scalar field, we meet condition i). This regime has been discussed, for instance, in  \cite{BarberoG:2007qra,BarberoG.:2007rd,martin2011,Martin-Benito:2013jqa}. 

Let us now focus on condition ii). The determinant of the spatial metric $\tilde h(x)$ defined in \eqref{eq:deth} is (approximately) independent of the inhomogeneous fields if the exponent in that expression, namely,
\be\label{eq:lin-reg}
X(x) = -\frac{\xi_1^2}{4a_2a_3}+\frac{2}{a_1p_{a_1}}\left[\frac{a_2p_{a_2}-a_2p_{a_2}}{2\sqrt{a_2a_3}}\xi_1-P_{\phi_0}\phi_1 -\int_0^z \,d\tilde z(\pi_{\xi_1} \xi_1'+P_{\phi_1}\phi_1')\right],
\ee
is constant or near to zero. One can check explicitly that $X(x)$ is not constant. Hence, we will consider those situations in which it approaches zero. This can be achieved dynamically, for instance, in an initially expanding spacetime with the scale factors reaching sufficiently large values such that they dilute the inhomogeneities in this expression (inhomogeneities themselves will dilute during expansion in the linear regime as we will see below). In this case $\tilde h(x)\simeq \tilde h=(a_1a_2a_3)^2$  becomes independent of the inhomogeneities. We will assume condition ii) is always satisfied in the remainder of this manuscript, and discuss its consequences.

The reduced scalar constraint can be approximated by
\bea\nonumber
&&\underline S_0^{(2)}\simeq \tilde{\underline S}_0^{(2)}=L_xL_yL_z \bigg[\frac{\kappa}{4}\bigg(a_1^2p_{a_1}^2+a_2^2p_{a_2}^2+a_3^2p_{a_3}^2-2a_1p_{a_1}a_2p_{a_2}-2a_1p_{a_1}a_3p_{a_3}-2a_2p_{a_2}a_3p_{a_3}\bigg)\\\nonumber
&&+\frac{P^2_{\phi_0}}{2}+\frac{1}{2}(a_1a_2a_3)^2M^2\phi_0^2\bigg]+L_yL_z \int_0^{L_x}dx\bigg[\kappa \bigg(\frac{a_1^2p_{a_1}^2}{16a_2a_3}\xi_1^2+a_2a_3\pi_{\xi_1}^2\bigg)+\frac{a_2a_3}{4\kappa}(\xi_1')^2\\
&&+\frac{P^2_{\phi_1}}{2}+\frac{(a_2a_3)^2}{2}(\phi_1')^2+\frac{1}{2}(a_1a_2a_3)^2M^2\phi_1^2\bigg].\label{eq:red-hamconst-quad}
\ea 

Let us now expand in Fourier modes the inhomogeneities. Concretely,
\be
\xi_1(x)=\sum_{n\neq 0}\tilde \xi_1(k_n)e^{ik_n\,x},
\ee
where $k_n=\frac{2 n \pi}{L_x}$ with $n=\pm 1,\pm 2,\ldots,\pm \infty$, since the inhomogeneous fields satisfy $\xi_1(x=0)=\xi_1(x=L_x)$, and similarly for $\phi_1$ and the momenta. Moreover, they are real fields. This implies, for instance, that 
\be
{\tilde{\xi}_1}(k_n)^*={\tilde \xi_1}(-k_n),
\ee
where the asterisk means complex conjugation. We adopt the same Fourier expansion for all other inhomogeneities. The Poisson brackets of the Fourier modes are
\be
\{\tilde \xi_1(k_n),\tilde P_{\xi_1}(k_{n'})\}=\frac{1}{L_xL_yL_z}\delta_{k_n,-k_{n'}}, \quad \{\tilde \phi_1(k_n),\tilde P_{\phi_1}(k_{n'})\}=\frac{1}{L_xL_yL_z}\delta_{k_n,-k_{n'}}.
\ee

After Fourier expansion, the above reduced scalar constraint takes the form
\bea\nonumber
&&\tilde{\underline S}_0^{(2)} =L_xL_yL_z \bigg[\frac{\kappa}{4}\bigg(a_1^2p_{a_1}^2+a_2^2p_{a_2}^2+a_3^2p_{a_3}^2-2a_1p_{a_1}a_2p_{a_2}-2a_1p_{a_1}a_3p_{a_3}-2a_2p_{a_2}a_3p_{a_3}\bigg)\\\nonumber
&&+\frac{P^2_{\phi_0}}{2}+\frac{1}{2}(a_1a_2a_3)^2M^2\phi_0^2+\sum_{n\neq 0}\kappa \bigg(\frac{a_1^2p_{a_1}^2}{16a_2a_3}|\tilde \xi_1|^2+a_2a_3|\tilde \pi_{\xi_1}|^2\bigg)+\frac{a_2a_3}{4\kappa}k_n^2|\tilde\xi_1|^2\\
&&+\frac{|\tilde P_{\phi_1}|^2}{2}+\frac{(a_2a_3)^2}{2}k_n^2|\tilde\phi_1|^2+\frac{1}{2}(a_1a_2a_3)^2M^2|\tilde \phi_1|^2\bigg],\label{eq:red-hamconst-fourier}
\ea 
and the reduced diffeomorphism constraint
\bea\label{eq:red-diffb}
\underline{\tilde V_x}^{(2)} &=&L_xL_yL_z \sum_{n\neq 0}ik_n\bigg[\tilde\pi^*_{\xi_1} \tilde\xi_1+\tilde P^*_{\phi_1}\tilde\phi_1\bigg].
\ea 

Now, in order to solve the system, we still need to specify conditions that fix $\underline N_0^x$ and $\underline N_0$. For the latter, we only need to choose an internal clock variable. Here, the choice will be such that  $\underline N_0=1/\sqrt{\tilde h}$. For the gauge fixing of $N_0^x$ we can adopt the following strategy. Let us consider the gauge fixing condition $\Phi_{\tilde \phi_1}=\tilde \phi_1(k_1)-\alpha$ with $\alpha$ a constant complex number. Preservation of this gauge fixing condition yields
\be
\dot\Phi_{\phi_1}=0,\quad \to \quad N_0^x i k_1 \alpha +\underline N_0\tilde P_{\phi_1}(k_1) = 0.
\ee
This last expression allows us to solve 
\be
 N_0^x =\frac{i\underline N_0\tilde P_{\phi_1}(k_1)}{k_1 \alpha} .
\ee
Now, we find the conjugate momentum by solving the (global) diffeomorphism constraint as
\bea\label{eq:moment-cond}
&&\alpha^* \tilde P_{\phi_1}(k_1) - \tilde P_{\phi_1}^*(k_{1})\alpha=\bigg[\tilde\pi^*_{\xi_1}(k_{1}) \tilde\xi_1(k_{1})-\tilde\pi_{\xi_1}(k_1) \tilde\xi_1^*(k_1)\bigg] \\\nonumber
&&+\sum_{n\neq 0,1}n\bigg[\tilde\pi_{\xi_1}^*(k_n) \tilde\xi_1(k_n)-\tilde\pi_{\xi_1}(k_n) \tilde\xi^*_1(k_n)+\tilde P_{\phi_1}^*(k_n)\tilde\phi_1(k_n)-\tilde P_{\phi_1}(k_n)\tilde\phi^*_1(k_n)\bigg]. 
\ea
The right hand side is fixed by the choice of the initial state of the inhomogeneities. Actually, one can see that
\be\label{eq:Ok}
O_{\tilde \xi_1}(k_n)=i\left(\tilde\pi_{\xi_1}^*(k_n) \tilde\xi_1(k_n)-\tilde\pi_{\xi_1}(k_n) \tilde\xi^*_1(k_n)\right),\quad O_{\tilde \phi_1}(k_n)=i\left(\tilde P_{\phi_1}^*(k_n) \tilde\phi_1(k_n)-\tilde P_{\phi_1}(k_n) \tilde\phi^*_1(k_n)\right),
\ee
for all $n$, are real constants of motion (their physical interpretation will be explained below). Therefore, if one chooses the right hand side of Eq. \eqref{eq:moment-cond} to be equal zero, we get $\alpha^* \tilde P_{\phi_1}(k_1) - \tilde P_{\phi_1}^*(k_{1})\alpha=0$ which implies $\tilde P_{\phi_1}(k_1)=0$ (provided $\alpha$ is not a real or pure imaginary number) and consequently $ N_0^x =0$, as we wish.

Hence, for this gauge fixing condition, the dynamics of the final reduced model is determined by the scalar constraint \eqref{eq:red-hamconst-fourier} provided the initial data satisfies the following set of conditions
\be\label{eq:total-P}
\sum_{n\neq 0}n\bigg[O_{\tilde \xi_1}(k_n)+O_{\tilde \phi_1}(k_n)\bigg]=0,\quad O_{\tilde \phi_1}(k_1) = 0,\quad \tilde \phi_1(k_1) = \alpha\neq 0.
\ee
Note that these conditions are not too restrictive since they still leave infinitely many physical degrees of freedom available. The final reduced  Hamiltonian will be
\be\label{eq:red-totham-gfixed}
\underline{\cal H}^{(2)} = \underline N_0\tilde{\underline S}_0^{(2)},
\ee
with $\tilde{\underline S}_0^{(2)}$ given in Eq. \eqref{eq:red-hamconst-fourier} but setting $\tilde P_{\phi_1}(k_1)=0$ and $\phi_1(k_1)=\alpha$ in the mode sum (recalling that $\alpha\in\mathbb{C}$ can be chosen to be a complex number with a negligibly small modulus). 

With all this in mind, one can compute the equations of motion of the system (see Appendix \ref{app:eom}) for $\underline N_0 = 1/\sqrt{\tilde h}$. On the one hand, on solutions, $\tilde{\underline S}_0^{(2)}=0$. On the other hand, the equations of motion of the directional scale factors can be combined with this last equation into an effective Friedmann equation for anisotropic spacetimes. Concretely, one can see that
\be\label{eq:aniso-fried-inhom}
H^2+\frac{(H_2+H_3)^{2}}{48  a_2 a_3}\sum_{n\neq 0}|\tilde{\xi}_1|^2=\frac{\kappa}{3}\left(\rho_{\phi_0}+\rho_{\phi_1}+\rho_{\xi_1}\right)+\frac{\sigma^2}{6},
\ee
where $H=\frac{1}{3}(H_1+H_2+H_3)$ is the mean Hubble parameter, and $H_i=\frac{\dot a_i}{a_i}$ are the directional Hubble parameters. Besides, we define  $\sigma^2=(H_1-H)^2+(H_2-H)^2+(H_3-H)^2$, and 
\bea
\rho_{\phi_0}&=&\frac{1}{2 a_1^{2} a_2^{2} a_3^{2}}\tilde{P}_{\phi_0}^2+\frac{1}{2}M^2\phi_0^2\\
\rho_{\phi_1}&=&\frac{1}{2 a_1^{2}}k_1^2|\alpha|^2+\frac{1}{2}M^2|\alpha|^2+\sum_{n\neq 0,1}\left(\frac{|\tilde{P}_{\phi_1}|^2}{2 a_1^{2} a_2^{2} a_3^{2}}+\frac{1}{2 a_1^{2}}k_n^2|\tilde{\phi}_1|^2+\frac{1}{2}M^2|\tilde{\phi}_1|^2\right),\\
\rho_{\xi_1}&=&\sum_{k_n\neq 0}\left(\frac{\kappa}{a_1^{2} a_2 a_3}|\tilde{\pi}_{\xi_1}|^2+\frac{1}{4 \kappa a_1^{2} a_2 a_3}k_n^2|\tilde{\xi}_1|^2\right),
\ea
can be interpreted as the total energy densities of the homogeneous and inhomogeneous contributions of the inflaton and the inhomogeneous geometrical (or gravitational wave) degrees of freedom, respectively. Note that they are all positive definite. Besides, it is interesting to note that the mode $\tilde\phi_1(k_1)=\alpha$ can contribute at late times (before reheating) as an effective positive cosmological constant $\rho_{\Lambda_{eff}}\simeq\frac{1}{2}M^2|\alpha|^2$. However, its magnitude is specified once and for all in the gauge fixing process (where there is no preferred choice). Moreover, other gauge fixings, like time-dependent ones, would not necessarily leave remnants in the Friedmann equation that behave as an effective cosmological constant.

All other inhomogeneous modes satisfy the second order, ordinary differential equations 
\bea\label{eq:xi1}
&&\ddot{\tilde{\xi}}_1(k_n)+H_1\dot{\tilde{\xi}}_1(k_n)+\frac{k_n^2}{a_1^2}\tilde\xi_{1}(k_n)+\frac{\kappa^2 p_{a_1}^2}{4a_2^2a_3^2}\tilde\xi_{1}(k_n)=0,\\\label{eq:phi1}
&& \ddot{\tilde{\phi}}_1(k_n) +3H\dot{\tilde{\phi}}_1(k_n)+\frac{k_n^2}{a_1^2}\tilde{\phi_1}(k_n)+M^2\tilde{\phi_1}(k_n)=0,
\ea
with real coefficients. Therefore, given a complex solution to these equations, the complex conjugate will be linearly independent. This can be rigorously formulated in terms of the Klein-Gordon inner product associated to each tensor and scalar mode equations. For instance, for the tensor modes, given two solutions ${}^{(a)}u$ and ${}^{(b)}u$ to the equations of motion, which share the same homogeneous trajectory given by $a_1(t)$, $a_2(t)$ and $a_3(t)$, it is defined as
\be\label{eq:ip-xi}
\langle {}^{(a)}u(k_n,t),{}^{(b)}u(k_n,t)\rangle = i L_xL_yL_z\left[{}^{(a)}\tilde{\xi}_1^*(k_n,t){}^{(b)}\tilde{\pi}_{\xi_1}(k_n,t)-{}^{(a)}\tilde{\pi}^*_{\xi_1}(k_n,t){}^{(b)}\tilde{\xi}_1(k_n,t)\right],
\ee
and similarly for scalar modes $\tilde{\phi}_1$, i.e.
\be\label{eq:ip-phi}
\langle {}^{(a)}v(k_n,t),{}^{(b)}v(k_n,t)\rangle = i L_xL_yL_z\left[{}^{(a)}\tilde{\phi}_1^*(k_n,t){}^{(b)}\tilde{P}_{\phi_1}(k_n,t)-{}^{(a)}\tilde{P}^*_{\phi_1}(k_n,t){}^{(b)}\tilde{\phi}_1(k_n,t)\right].
\ee
This inner product is conserved with respect to time evolution. But we should note that this is true if and only if the two solutions $(a)$ and $(b)$ produce the same reaction on the homogeneous sector. Otherwise, the above inner products are not preserved in time and therefore \eqref{eq:ip-xi} and \eqref{eq:ip-phi} are meaningless.  Hence, in summary, it only makes sense to consider solutions within the equivalence class that produce exactly the same reaction on the homogeneous sector.  

Thus, we can express the Fourier modes of both inhomogeneities as
\bea\nonumber
&&\tilde \xi_1(k_n,t) = a_{\tilde \xi_1,u}(k_n)u_{\tilde \xi_1}(k_n,t)+a^*_{\tilde \xi_1,u}(-k_n)u^*_{\tilde \xi_1}(-k_n,t),\\\label{eq:Fmodes}
&&\tilde \phi_1(k_n,t) = a_{\tilde \phi_1,v}(k_n)v_{\tilde \phi_1}(k_n,t)+a^*_{\tilde \phi_1,v}(-k_n)v^*_{\tilde \phi_1}(-k_n,t),
\ea
with $u_{\tilde \xi_1}(k_n,t)$ and $v_{\tilde \phi_1}(k_n,t)$ two complex solutions to the equations of motion \eqref{eq:xi1} and \eqref{eq:phi1} with unit norm  with respect to the inner products \eqref{eq:ip-xi} and \eqref{eq:ip-phi}, respectively. Besides, $a_{\tilde \xi_1,u}(k_n)$ and $a_{\tilde \phi_1,v}(k_n)$ are complex constants of motion that determine the initial state of the Fourier modes, and therefore, of the inhomogeneities. They satisfy the Poisson algebra
\be
\{a_{\tilde \xi_1,u}^*(k_n),a_{\tilde \xi_1,u}(k_{n'})\}=i\delta_{k_n,k_{n'}},\quad \{a_{\tilde \phi_1,v}^*(k_n),a_{\tilde \phi_1,v}(k_{n'})\}=i\delta_{k_n,k_{n'}}.
\ee

Moreover, one can easily see that the observables $O_{\tilde \xi_1}(k_n)$ and $O_{\tilde \phi_1}(k_n)$ can be expressed in terms of $a_{\tilde \xi_1,u}(k_n)$ and $a_{\tilde \phi_1,v}(k_n)$ as
\be
O_{\tilde \xi_1}(k_n)=|a_{\tilde \xi_1,u}(k_n)|^2-|a_{\tilde \xi_1,u}(-k_n)|^2,\quad O_{\tilde \phi_1}(k_n)=|a_{\tilde \phi_1,v}(k_n)|^2-|a_{\tilde \phi_1,v}(-k_n)|^2.
\ee
Then, one can interpret $a_{\tilde \xi_1,u}(k_n)$ and $a_{\tilde \xi_1,u}^*(k_n)$ as a complex amplitude of positive and negative frequency modes with respect to $(u_{\tilde\xi_1},u^*_{\tilde\xi_1})$, with $O_{\tilde \xi_1}(k_n)$ measuring the difference between the two. This interpretation also applies for the modes of the scalar field $\phi_1$. Finally, it is worth mentioning that Eq. \eqref{eq:total-P} is naturally interpreted as a conservation law of the total momentum carried by the positive and negative frequency modes of the inhomogeneities.

This interpretation is valid for a fixed choice of the basis of solutions. But, given two different sets of solutions, it is very easy to construct the Bogoliubov transformation relating them. Let us consider a new basis of complex solutions $\tilde u_{\tilde \xi_1}(k_n,t)$. Then, we can express it as
\be
\tilde u_{\tilde \xi_1}(k_n,t) = \alpha_{\tilde u,u}(k_n) \,u_{\tilde \xi_1}(k_n,t)+\beta_{\tilde u,u}(k_n) \, u^*_{\tilde \xi_1}(k_n,t),
\ee
with the Bogoliubov coefficients defined as
\be
\alpha_{\tilde u,u}(k_n) = \langle u_{\tilde \xi_1}(k_n,t),\tilde u_{\tilde \xi_1}(k_n,t)\rangle,\quad \beta_{\tilde u,u}(k_n) = -\langle u^*_{\tilde \xi_1}(k_n,t),\tilde u_{\tilde \xi_1}(k_n,t)\rangle.
\ee
They satisfy $|\alpha_{\tilde u,u}(k_n)|^2-|\beta_{\tilde u,u}(k_n)|^2=1$. Then, one can also see that 
\be
a_{\tilde \xi_1,u}(k_n) = \alpha_{\tilde u,u}(k_n) \, a_{\tilde \xi_1,\tilde u}(k_n)+\beta_{\tilde u,u}^*(k_n)  \,a^*_{\tilde \xi_1,\tilde u}(k_n).
\ee
It is not difficult to realized that under these transformations the modes in Eq. \eqref{eq:Fmodes} (and their time derivatives) remain invariant. Therefore, the reaction of inhomogeneities on the homogeneous sector will also remain invariant. But one must keep in mind that the constants of motion $a_{\tilde \xi_1,u}(k_n) \neq a_{\tilde \xi_1,\tilde u}(k_n)$, unless the Bogoliubov transformation is the identity. Hence, there is no universal physical interpretation of the constants of motion $a_{\tilde \xi_1,u}(k_n)$ and $a^*_{\tilde \xi_1,u}(k_n)$. The previous discussion for the inhomogeneous modes of $\xi_1$ also applies to the ones of the scalar field $\phi_1$.

One question we can ask is how large is this family of solutions for which the inner products \eqref{eq:ip-xi} and \eqref{eq:ip-phi} are preserved. Actually, from the viewpoint of the choice of complex basis of solutions, there are infinitely many possibilities available. However, we should note that any of those choices must leave the homogeneous sector of the theory invariant, namely, which implies a strong restriction in the inhomogeneous sector since the absolute values of the Fourier modes $\tilde\xi_1(k_,t)$, $\tilde\phi_1(k_,t)$, and the one of their time derivatives, must remain also invariant. We still have the freedom of adding a global mode-by-mode complex phase. However, such a phase can be absorbed as a trivial Bogoliubov transformation. From this perspective, the freedom is actually quite small since the choice of basis leaves the solution space of the homogeneous and inhomogeneous sectors invariant. However, it is worth mentioning that, although this is the case in the classical system, if one quantizes the inhomogeneities, promoting the annihilation and creation variables to quantum operators, one could expect different results. In this case, although the semiclassical backreaction of inhomogeneities will require regularization and renormalization in order to give sensible physical results, the choice of basis of solutions will determine the Fock vacuum.\footnote{A choice of basis of solutions is tantamount to select a particular Fock vacuum. In this sense, two choices of basis yield vacuum states within the same Fock Hilbert space if $|\beta_{\tilde u,u}^*(k_n)|^2$ is summable in $k_n$. Otherwise, they will define different (i.e. unitarily inequivalent) Fock representations. It is interesting to note that there are criteria for the selection of vacuum state that can be easily adopted in these Gowdy scenarios. See, for instance,  Refs. \cite{deBlas:2016puz,ElizagaNavascues:2020fai,ElizagaNavascues:2018bgp,ElizagaNavascues:2019itm,Martin-Benito:2021szh,Martin-Benito:2021ulw} and references therein.} Therefore, one should expect different results for the semiclassical backreaction of inhomogeneities. 

Let us also mention that we have also analyzed in some detail the isotropic limit of the homogeneous sector of the model (see Appendix \ref{app:iso-sys}), reaching similar conclusions. In particular, we derive the corresponding Friedmann equation  \eqref{eq:fried-inhom}. We note that tensor inhomogeneities are constrained by Eq. \eqref{eq:collapse-cond}. If the scale factor is expanding and condition  \eqref{eq:collapse-cond} is fulfilled, all tensor inhomogeneities will not cause the spacetime to recollapse at later times since 
\be
\frac{1}{12a^2}\sum_{n\neq 0}|\tilde\xi_1(k_n)|^2,
\ee
always decreases with the expansion. This is the case because of the positivity of the coefficients in the equations of motion \eqref{eq:xi1_iso} of the modes. More concretely, positivity of the last term multiplying the mode will never induce tachyonic instabilities (i.e. inhomogeneities do not blow up) while the one multiplying the velocity is a positive friction term that will cause a damping in the amplitude of the modes. This arguments also apply for the scalar modes. 

It is also interesting to compare the Friedmann equations in expressions \eqref{eq:aniso-fried-inhom} and \eqref{eq:fried-inhom}. At first sight one can think that there is a disagreement in the isotropic limit since their left hand sides (that contain information about the Hubble parameters) disagree in one sign. However, one must note that we define the isotropic limit on phase space and that the quantity $\sigma^2$ does not vanishes in presence of tensor inhomogeneities in this isotropic limit. Indeed, one could interpret the extra factor on the left hand side of Eq. \eqref{eq:fried-inhom} as a spacetime shear sourced by inhomogeneities (once it has been moved to the right hand side). Concretely, given Eq. \eqref{eq:red-ds2} and setting $\tilde h=(a_1a_2a_3)^2$ and $M=0$ (for simplicity), one can see that $\tau_a=\nabla_a \tau$, being $\tau$ cosmic time defines a congruence of geodesics. Then, the traceless part of $\nabla_a\tau_b$ determines the spacetime shear. In the isotropic sector $a_1=a_2=a_3$, the shear does not vanish unless $\xi_1(x)=0$. On the other hand, in the homogeneous limit $\xi_1(x)=0=\phi_1(x)$ (and their momenta), the quantity $\sigma^2$ agrees with the spacetime shear of Bianchi I cosmology, which vanishes in the isotropic limit. 

In summary, this will be the setting that will allow us to evolve the system from an inhomogeneous pre-inflationary regime, where the potential of the scalar field is negligible, to another regime where the potential can dominate the evolution and check if it meets the requirements to start a phase of nearly exponential expansion, namely, the onset of inflation. At this time, one can check if condition \eqref{eq:lin-reg} is met, since the inflaton potential becomes relevant. In the affirmative case, the inhomogeneities will satisfy a linear dynamics and the discussion above will still be valid. Otherwise, the lack of linearity will require a deeper analysis since the Klein-Gordon inner products as written in Eqs. \eqref{eq:xi1} and \eqref{eq:phi1} will not be time-independent.

\section{Conclussions}\label{sec:concl}

We have shown that polarized Gowdy cosmologies coupled to a massive scalar field admit a phase space with a natural splitting in homogeneous and inhomogeneous sectors. When inhomogeneities vanish, the homogeneous sector corresponds to a Bianchi I model coupled to a homogeneous massive scalar field, and is therefore of physical interest \cite{Pereira:2007,Pitrou:2008,Agullo:2020-lett}. On the other hand, inhomogeneities of both geometry and matter satisfy linear equations of motion either when the mass of the scalar field is negligible or if the spacetime has expanded enough such that the spatial volume element is nearly independent of the inhomogeneities. In a universe dominated by the kinetic energy of the inflation field at early times one reaches this situation.  Assuming that the spacetime is initially expanding, inhomogeneities will in general dilute as the scale factors expand. Concretely, preliminary simulations show that at initial times, where the kinetic energy of the homogeneous mode of the scalar field dominates over the inhomogeneous energy densities $\rho_{\phi_1}$ and $\rho_{\xi_1}$, the latter dilute at a slightly slower rate than the former, and they can eventually dominate. However, the potential energy of the homogeneous mode of the scalar field, which is nearly constant, rapidly takes over and becomes the dominant contribution. Afterwards, inflation begins, and all other inhomogeneous contributions to the Friedmann equation dilute or become subdominant. Besides, by the time the model reaches the onset of inflation, where the potential of the inflaton dominates, the spatial volume is nearly independent of the inhomogeneities. Then, the evolution of the latter is nearly linear there and later on, until the end of inflation. It is also worth to mention that, if anisotropies are important at early times, there exists the possibility where the scale factor $a_1$ could begin contracting despite the mean scale factor $a$ is expanding. Hence, tensor inhomogeneities can actually grow. This special situations will be studied in detail in a forthcoming contribution. Besides, we have noted that in the reduction and gauge fixing process, one of the large scale inhomogeneous modes of the scalar field remains frozen and can contribute at late times as an effective, small and positive cosmological constant, before this inflaton field decays during reheating. But we should remind that this is a consequence of a gauge fixing where we choose its magnitude to be small by convenience. Other choices could might not yield to such a contribution. Whether it is likely or not, it is still a possible mechanism for dark energy production at very early times that deserves to be explored in the future.

Within this linear approximation we show that one can follow the standard treatment for linear field theories. Namely, we identify a suitable basis of orthonormal complex solutions to the equations of motion. They have positive norm with respect to the Klein-Gordon inner product. We noted that, given two complex inhomogeneous solutions, the inner product is time-independent if and only the reaction of the inhomogeneous solutions on the homogeneous sector is the same at all times. We then expand the Fourier modes of the inhomogeneities as linear combinations of these complex solutions. The coefficients represent creation and annihilation variables. They are constants of motion that characterize the initial state of the inhomogeneous sector and represent the complex amplitude of the positive and negative frequency modes. We also show that it is possible to choose different basis of solutions for the Fourier modes, provided the latter remain invariant. This implies that one needs to compensate the change accordingly in the creation and annihilation variables (constants of motion) via a well-known Bogoliubov transformation. Therefore, the physical meaning of these creation and annihilation variables is always relative to the choice of orthonormal basis of solutions. The reason for this compensation is to guaranty that the reaction of the inhmogeneities on the homogeneous sector is preserved dynamically, as well as the inner products \eqref{eq:ip-xi} and \eqref{eq:ip-phi}. We have also discussed that once a given trajectory of the homogeneous sector is chosen, the only freedom left in the inhomogeneous sector is in the multiplication by a set of mode-by-mode time-independent complex phases, which can actually be absorbed in a Bogoliubov transformation. Hence, the freedom that is left is quite small. Nevertheless, in a quantum treatment, different choices of Fock vacua for the inhomogeneities are expected to contribute with different backreaction. This is however out of the scope of this manuscript.

The results obtained in this paper provide a theoretical framework for the numerical study of the classical dynamics of this setting. For instance, it will be possible to analyze the robustness of inflation in presence of nonperturbative inhomogeneities in a kinetically dominated early universes. It will also allow us to eventually investigate a Fock quantization for the inhomogeneities and its physical consequences within the paradigm of inflation. For instance, a very anisotropic and inhomogeneous pre-inflationary phase is expected to drive scalar and tensor perturbations into an excited state with respect to the Bunch-Davies vacuum, which can leave imprints in the CMB. This will be however a matter if future research.

\acknowledgements

We acknowledge G. Garc\'ia-Moreno and B. Elizaga Navascués for stimulating discussions. This work is supported by the Spanish Government through the projects FIS2017-86497-C2-2-P, PID2019-105943GB-I00 (with FEDER contribution), and the "Operative Program FEDER2014-2020 Junta de Andaluc\'ia-Consejer\'ia de Econom\'ia y Conocimiento" under project E-FQM-262-UGR18 by Universidad de Granada.

\appendix

\section{Symmetry reduction}\label{app:sym-gauge-red}

In this appendix we explain the classical reduction, and subsequently, judicious gauge fixing conditions already introduced in the literature dedicated to polarized Gowdy models \cite{gowdy1971gravitational,gowdy1974vacuum}. Let us introduce two commuting, hypersurface orthogonal, spacelike Killing vector fields with compact orbits, $(\partial_y)^i$ and $(\partial_z)^i$. The Killing fields allow us to eliminate the dependence of metric (and all other phase space variables) on the coordinates adapted to the Killing fields, in this case $(y,z)$. Once this symmetry is imposed, we proceed by solving the momentum constraints in the $y$ and $z$ directions. For this purpose, we introduce the gauge fixing conditions $h_{xy}=0=h_{xz}$. Their preservation in time fix the shift components $N_y= 0= N_z$, as well as the corresponding momenta of $h_{xy}$ and $h_{xz}$ can be obtained by solving the corresponding diffeomorphism constraints. In addition, when the Killing fields are orthogonal, as it is the case here of polarized Gowdy model, an additional local degree of freedom is eliminated, namely $h_{yz}=0$ (if they are not orthogonal, one would obtain the unpolarized Gowdy model instead). Let us note that this is a physical restriction and not a gauge fixing. The condition $\dot h_{yz}=0$ fixes the conjugated momenta, i.e. we do not need to further solve any constraint neither fix a Lagrange multiplier. The final reduced spatial metric (and conjugate momentum) turns out to be diagonal. Besides, all reduced phase space fields only depend on $x$, the coordinate in the direction of the inhomogeneities.

After this symmetry reduction and gauge fixing, the spatial metric and its conjugate momentum take the form
\be \label{back} \tilde h_{ij}={\rm diag}(g_{1}(x),g_{2}(x),g_{3}(x)) \, ,  \hspace{1cm} \tilde p^{ij}={\rm diag}\left(p_{1}(x),p_{2}(x),p_{3}(x)\right)  \, .\ee
with Poisson brackets\footnote{The symplectic structure, since it involves a volume integral over the two Killing coordinates, will result in Poisson brackets where two of the original Dirac delta functions will be replaced by the factor $(L_yL_z)^{-1}$.}
\be \label{PB-grav} 
\{ g_{I}(x),p^{J}(x')\}=\frac{1}{L_yL_z}\delta_{I}^J\delta(x-x')\, ,\ee
with $I,J=1,2,3$. Similarly, the reduced matter degrees of freedom satisfy
\be \label{PB-phi} 
\{ \phi(x),P_{\phi}(x')\}=\frac{1}{L_yL_z}\delta(x-x')\, .\ee

The scalar constraint, after reduction and integrating out the Killing coordinates, turns out to be
\bea\nonumber
\tilde S(x) &=&L_yL_z \bigg[\frac{2\kappa}{\sqrt{\tilde h}} \bigg(\sum_{I=1}^3g_I^2p_I^2-\frac{1}{2}\left(\sum_{I=1}^3g_Ip_I\right)^2\bigg)-\frac{\sqrt{\tilde h}}{2\kappa}\bigg(\frac{g_1'g_2'}{2g_1^2g_2}+\frac{g_1'g_3'}{2g_1^2g_3}-\frac{g_2'g_3'}{2g_1g_2g_3}\\
&&+\frac{(g_2')^2}{2g_1g_2^2}+\frac{(g_3')^2}{2g_1g_3^2}-\frac{g_2''}{g_1g_2}-\frac{g_3''}{g_1g_3}\bigg)+\frac{P^2_{\phi}}{2\sqrt{\tilde h}}+\sqrt{\tilde h}V(\phi)+\frac{\sqrt{\tilde h}}{2g_1}(\phi')^2\bigg].
\ea 
Similarly, the remaining diffeomorphism constraint in the $x$ direction is
\bea
\tilde V_x(x) &=&L_yL_z \bigg[\bigg(\sum_{I=1}^3g_I'p_I\bigg)-2(g_1p_1)'+\phi'P_{\phi}\bigg].
\ea 
The total reduced Hamiltonian takes the form 
\be
\tilde{\cal H} = \int_0^{L_x}dx \left(\tilde N(x)\tilde S(x)+N^x(x)\tilde V_x(x)\right).
\ee

We will introduce a canonical transformation that will render our model in a more familiar form. This is a judicious transformation replaces our reduced phase space variables by new configuration variables 
\be\label{eq:gi}
g_1 = e^{\gamma-\frac{\xi}{\sqrt{\tau}}-\frac{\xi^2}{4\tau}},\quad g_2 = \tau^2 e^{-\frac{\xi}{\sqrt{\tau}}}, \quad g_3 = e^{\frac{\xi}{\sqrt{\tau}}},
\ee  
and momenta
\bea
p_1 &=& e^{-\gamma+\frac{\xi}{\sqrt{\tau}}+\frac{\xi^2}{4\tau}} \pi_\gamma,\\\nonumber
p_2 &=& e^{\frac{\xi}{\sqrt{\tau}}}\frac{\left(\xi\pi_\xi+2\tau\pi_\tau\right)}{4\tau^2},\\\nonumber
p_3 &=&  e^{-\frac{\xi}{\sqrt{\tau}}}\left(\frac{\xi\pi_\gamma}{2\sqrt{\tau}}+\pi_\gamma+ \frac{\xi\pi_\xi}{4}+\pi_\xi\sqrt{\tau}+\frac{\tau\pi_\tau}{2}\right).
\ea
One can verify that the new variables satisfy the Poisson brackets
\be \label{PB-grav-tau-gamma} 
\{ \tau(x),\pi_\tau(x')\}=\frac{1}{L_yL_z}\delta(x-x')\, ,\quad \{ \gamma(x),\pi_\gamma(x')\}=\frac{1}{L_yL_z}\delta(x-x')\, ,\quad \{ \xi(x),\pi_\xi(x')\}=\frac{1}{L_yL_z}\delta(x-x')\, .\ee

The scalar constraint, after absorbing a volume element $\sqrt{h}$ (due to a densitization of the lapse function as $\underline{N}=\frac{\tilde N}{\sqrt{h}}$) with $h=\tau^2 e^{\gamma-\frac{\xi}{\sqrt{\tau}}-\frac{\xi^2}{4\tau}}$, takes the form
\bea\nonumber
\underline S(x) &=&L_yL_z \bigg[\kappa \bigg(\frac{\xi^2\pi_\gamma^2}{4\tau}+\pi_\xi^2\tau-2\pi_\gamma\pi_\tau\tau\bigg)+\frac{1}{\kappa}\bigg(\frac{\tau(\xi')^2}{4}-\frac{\tau\tau'\gamma'}{2}-\frac{\xi^2(\tau')^2}{16\tau}+\tau\tau''\bigg)\\
&&+\frac{P^2_{\phi}}{2}+\frac{\tau^2}{2}(\phi')^2+\tau^2e^{\gamma-\frac{\xi}{\sqrt{\tau}}-\frac{\xi^2}{4\tau}}V(\phi)\bigg],
\ea 
Similarly, the diffeomorphism constraint is
\bea
\underline V_x(x) &=&L_yL_z \bigg[\pi_\tau\tau' +\pi_\gamma\gamma'+\pi_\xi \xi'-2\pi_\gamma'+\phi'P_{\phi}\bigg].
\ea 
The total reduced Hamiltonian is then 
\be
\underline{\cal H} = \int_0^{L_x}dx \left(\underline N(x)\underline S(x)+N^x(x)\underline V_x(x)\right).
\ee
The reduced spacetime metric is
\be
ds^2=-(\tau^2 \underline N^2-(N^x)^2)\frac{h}{\tau^2}dt^2+2\frac{h}{\tau^2}N^xdxdt+\frac{h}{\tau^2}dx^2+\tau^2 e^{-\frac{\xi}{\sqrt{\tau}}}dy^2+e^{\frac{\xi}{\sqrt{\tau}}}dz^2.
\ee

This canonical transformation will help us to split our inhomogeneous spacetime, after suitable gauge fixings, into homogeneous and inhomogeneous sectors. These choices are motivated by the fact that inhomogeneities satisfying linear (partial) differential equations in the absence of the potential $V(\phi)$.

\section{Gauge fixing and mode expansion}\label{sec:gauge-fix}

In this section we will introduce suitable partial gauge fixing conditions, starting with  $\pi_\gamma'=0$, and afterwards, with $\tau'=0$. These gauge fixing conditions have been already discussed in the literature for polarized Gowdy vacuum models. See for instance Ref. \cite{MenaMarugan:1998}. For the nonvacuum case see \cite{BarberoG:2007qra}. Finally, we will discuss a mode expansion for the reduced theory in order to characterize the homogeneous and inhomogeneous sectors. 

\subsection{Gauge fixing $\gamma$-sector}

In order to implement the gauge fixing condition $\pi_\gamma'=0$, it will be convenient to split the $\gamma$-sector into homogeneous and inhomogeneous degrees of freedom. Concretely, we express $\gamma(x)=\gamma_0+\gamma_1(x)$, such that 
\be\label{eq:gamma0}
 \gamma_0 = \frac{1}{L_x}\int_0^{L_x}dx\,\gamma(x)\,,\quad \int_0^{L_x}dx\,\gamma_1(x) = 0. 
\ee
This splitting also applies to the momentum variable. It yields the following Poisson bracket structure in the $\gamma$-sector 
\be \label{PB-grav-gamma}
\{ \gamma_0,\pi_{\gamma_0}\}=\frac{1}{L_xL_yL_z}\, ,\quad \{ \gamma_1(x),\pi_{\gamma_1}(x')\}=\frac{1}{L_yL_z}\delta(x-x')\, .
\ee
We then impose the gauge fixing condition $\Phi_{\gamma_1} = \pi_{\gamma_1}'$. Concretely, its dynamical preservation, namely $\dot \Phi_{\gamma_1} = 0$, yields a condition for the shift 
\be
[\underline N^x(x)]'' = \frac{1}{2\kappa \pi_{\gamma_0}}\left[(\underline N\tau\tau')'+\underline N\kappa \tau^2e^{\gamma_0+\tilde\gamma_1-\frac{\xi}{\sqrt{\tau}}-\frac{\xi^2}{4\tau}}V(\phi)\right]',
\ee
where 
\be
\tilde\gamma_1(x) = -\frac{1}{\pi_{\gamma_0}}\int_0^x \,d\tilde x(\pi_\tau\tau'+\pi_\xi \xi'+\phi'P_{\phi}),
\ee
is obtained after solving the corresponding diffeomorphism constraint. This gauge fixing condition fixes the shift function, except for a homogeneous Lagrange multiplier that we denote by $\underline N^x_0$. The inhomogeneous part of the shift will be $\underline N^x_1$. Then, 
\be
[\underline N^x(x)]' = \frac{1}{2\kappa \pi_{\gamma_0}}\left[(\underline N\tau\tau')'+\underline N\kappa \tau^2e^{\gamma_0+\tilde\gamma_1-\frac{\xi}{\sqrt{\tau}}-\frac{\xi^2}{4\tau}}V(\phi)-\frac{1}{L_x}\int_0^{L_x} \,d\tilde x\underline N\kappa \tau^2e^{\gamma_0+\tilde\gamma_1-\frac{\xi}{\sqrt{\tau}}-\frac{\xi^2}{4\tau}}V(\phi)\right],
\ee
with the last addend being a constant of integration (independent of $x$) that is fixed by the condition that the right hand side of this equation is a pure inhomogeneous function, as the left hand side.

The reduced phase space now does not depend on $(\gamma_1,\pi_{\gamma_1})$. The reduced constraints now read
\bea\nonumber
\underline S^{(1)}(x) &=&L_yL_z \bigg[\kappa \bigg(\frac{\xi^2\pi_{\gamma_0}^2}{4\tau}+\pi_\xi^2\tau-2\pi_{\gamma_0}\pi_\tau\tau\bigg)+\frac{1}{\kappa}\bigg(\frac{\tau(\xi')^2}{4}+\frac{\tau\tau'}{2\pi_{\gamma_0}}\left(\pi_\tau\tau'+\pi_\xi \xi'+\phi'P_{\phi}\right)\\
&&-\frac{\xi^2(\tau')^2}{16\tau}+\tau\tau''\bigg)+\frac{P^2_{\phi}}{2}+\frac{\tau^2}{2}(\phi')^2+\tau^2e^{\gamma_0+\tilde\gamma_1-\frac{\xi}{\sqrt{\tau}}-\frac{\xi^2}{4\tau}}V(\phi)\bigg],
\ea 
which is still a local constraint,\footnote{It is local in the sense that it will remove one degree of freedom per point in the circle $S^1$. However, the volume element multiplying the potential $V(\phi)$ now depends on $\tilde \gamma_1(x)$ which is a nonlocal function of the reduced phase spaces variables, in the sense that it involves a spatial integral.} unlike the remaining diffeomorphism constraint, which takes the form
\bea
\underline V_x^{(1)} &=&L_yL_z \int_{0}^{L_x}\,dx\bigg[\pi_\tau\tau' +\pi_\xi \xi'+\phi'P_{\phi}\bigg].
\ea 
The total reduced Hamiltonian is then 
\be
\underline{\cal H}^{(1)} = \underline N_0^x\underline V_x^{(1)}+\int_0^{L_x}dx \underline N(x)\underline S^{(1)}(x).
\ee

\subsection{Gauge fixing $\tau$-sector}

In order to implement the next gauge fixing condition, we will also adopt a similar splitting of the $\tau$-sector into homogeneous and purely inhomogeneous fields. Namely,  $\tau(x)=\tau_0+\tau_1(x)$ and $\pi_\tau(x)=\pi_{\tau_0}+\pi_{\tau_1}(x)$. The new Poisson brackets are then  
\be \label{PB-grav-tau1}
\{ \tau_0,\pi_{\tau_0}\}=\frac{1}{L_xL_yL_z}\, ,\quad \{ \tau_1(x),\pi_{\tau_1}(x')\}=\frac{1}{L_yL_z}\delta(x-x')\, .
\ee
In addition, it is also convenient to split in the same way the densitized lapse function as $\underline N(x)=\underline N_0+\underline N_1(x)$. The new total Hamiltonian takes the form
\be
\underline{\cal H}^{(1)} = N_0^x\underline V_x^{(1)}+\underline N_0 \underline S_0^{(1)}+\int_0^{L_x}dx \underline N_1(x)\underline S_1^{(1)}(x),
\ee
now with
\bea\nonumber
\underline S_0^{(1)} &=&L_yL_z \int_0^{L_x}dx\bigg[\kappa \bigg(\frac{\xi^2\pi_{\gamma_0}^2}{4(\tau_0+\tau_1)}+\pi_\xi^2(\tau_0+\tau_1)-2\pi_{\gamma_0}\pi_{\tau_0}\tau_0\bigg)+\frac{1}{\kappa}\bigg(\frac{(\tau_0+\tau_1)(\xi')^2}{4}\\\nonumber
&&+\frac{(\tau_0+\tau_1)\tau_1'}{2\pi_{\gamma_0}}\left((\pi_{\tau_0}+\pi_{\tau_1})\tau_1'+\pi_\xi \xi'+\phi'P_{\phi}\right)-\frac{\xi^2(\tau_1')^2}{16(\tau_0+\tau_1)}+\tau_1\tau_1''\bigg)\\
&&+\frac{P^2_{\phi}}{2}+\frac{(\tau_0+\tau_1)^2}{2}(\phi')^2+(\tau_0+\tau_1)^2e^{\gamma_0+\tilde\gamma_1-\frac{\xi}{\sqrt{\tau_0+\tau_1}}-\frac{\xi^2}{4(\tau_0+\tau_1)}}V(\phi)\bigg],\\
\underline S^{(1)}(x) &=&L_yL_z \bigg[\kappa \bigg(\frac{\xi^2\pi_{\gamma_0}^2}{4(\tau_0+\tau_1)}+\pi_\xi^2(\tau_0+\tau_1)-2\pi_{\gamma_0}(\pi_{\tau_0}\tau_1+\tau_0\pi_{\tau_1}+\tau_1\pi_{\tau_1})\bigg)\\\nonumber
&&+\frac{1}{\kappa}\bigg(\frac{(\tau_0+\tau_1)(\xi')^2}{4}+\frac{(\tau_0+\tau_1)\tau_1'}{2\pi_{\gamma_0}}\left((\pi_{\tau_0}+\pi_{\tau_1})\tau_1'+\pi_\xi \xi'+\phi'P_{\phi}\right)-\frac{\xi^2(\tau_1')^2}{16(\tau_0+\tau_1)}\\\nonumber
&&+(\tau_0+\tau_1)\tau_1''\bigg)+\frac{P^2_{\phi}}{2}+\frac{(\tau_0+\tau_1)^2}{2}(\phi')^2+(\tau_0+\tau_1)^2e^{\gamma_0+\tilde\gamma_1-\frac{\xi}{\sqrt{(\tau_0+\tau_1)}}-\frac{\xi^2}{4(\tau_0+\tau_1)}}V(\phi)\bigg]
\ea 
 
We impose the new gauge fixing condition $\Phi_{\tau_1} = {\tau_1}$. Its dynamical preservation, i.e. $\dot \Phi_{\tau_1}=0$, yields the condition 
\be
-2\kappa \pi_{\gamma_0}\tau_0 \underline N_1(x) = 0\,,\quad \to \quad \underline N_1(x) = 0\,.
\ee
Now, solving $\pi_{\tau_1}$ using the constraint $\underline S^{(1)}(x)$, we get
\be
\pi_{\tau_1} = \frac{1}{2\pi_{\gamma_0}}\left(\pi_{\xi}^2+\frac{{\xi}^2\pi_{\gamma_0}^2}{4\tau_0^2}+\frac{P_\phi^2}{2\kappa \tau_0}+\frac{{\xi'}^2}{4\kappa^2}+\frac{\tau_0{\phi'}^2}{2\kappa}+\frac{1}{2\kappa}\tau_0e^{\gamma_0+\tilde\gamma_1-\frac{\xi}{\sqrt{\tau_0}}-\frac{\xi^2}{4\tau_0}}V(\phi)\right).
\ee
Note that this gauge fixing condition leaves the homogeneous contribution to the densitized lapse $\underline N_0$ undetermined. The reduced phase space now does not depend on $(\tau_1,\pi_{\tau_1})$. The remaining new reduced scalar constraint is 
\bea\nonumber
\underline S_0^{(2)} &=&L_yL_z \int_0^{L_x}dx\bigg[\kappa \bigg(\frac{\xi^2\pi_{\gamma_0}^2}{4\tau_0}+\pi_\xi^2\tau_0-2\pi_{\gamma_0}\pi_{\tau_0}\tau_0\bigg)+\frac{\tau_0(\xi')^2}{4\kappa}+\frac{P^2_{\phi}}{2}+\frac{\tau_0^2}{2}(\phi')^2\\
&&+\tau_0^2e^{\gamma_0+\bar\gamma_1-\frac{\xi}{\sqrt{\tau_0}}-\frac{\xi^2}{4\tau_0}}V(\phi)\bigg],
\ea 
where 
\be
\bar\gamma_1(x) = -\frac{1}{\pi_{\gamma_0}}\int_0^{x} \,d\tilde x(\pi_\xi \xi'+\phi'P_{\phi}).
\ee
On the other hand, the new reduced diffeomorphism constraint is given by
\bea
\underline V_x^{(2)} &=&L_yL_z \int_{0}^{L_x}\,dx\bigg[\pi_\xi \xi'+\phi'P_{\phi}\bigg].
\ea 
The total reduced Hamiltonian is then 
\be
\underline{\cal H}^{(2)} = N_0^x\underline V_x^{(2)}+\underline N_0\underline S^{(2)}.
\ee

Finally, let us split both $\phi(x)=\phi_0+\phi_1(x)$ and $\xi(x)=\xi_0+\xi_1(x)$, with $\phi_1(x)$ and $\xi_1(x)$ pure inhomogeneous fields (and similarly for their conjugate momenta). One can easily verify that the reduced spacetime line element is given by 
\be\label{eq:red-ds2-old}
ds^2=-(\tau_0^2 \underline N_0^2-(N_0^x+N_1^x)^2)\frac{\tilde h}{\tau_0^2}dt^2+2\frac{\tilde h}{\tau_0^2}(N^x_0+N^x_1)dxdt+\frac{\tilde h}{\tau_0^2}dx^2+\tau_0^2 e^{-\frac{\xi_0+\xi_1}{\sqrt{\tau_0}}}dy^2+e^{\frac{\xi_0+\xi_1}{\sqrt{\tau_0}}}dz^2.
\ee
where $\tilde h=\tau_0^2e^{\gamma_0-\frac{1}{\pi_{\gamma_0}}(\pi_{\xi_0} \xi_1+P_{\phi_0}\phi_1)-\frac{1}{\pi_{\gamma_0}}\int_0^z \,d\tilde z(\pi_{\xi_1} \xi_1'+P_{\phi_1}\phi_1')-\frac{\xi_0+\xi_1}{\sqrt{\tau_0}}-\frac{(\xi_0+\xi_1)^2}{4\tau_0}}$ 
and
\bea\label{eq:shift-red-old}\nonumber
&&[\underline N_1^x(x)]' = \frac{\underline N_0 }{4\pi_{\gamma_0}}\tilde h M^2(\phi_0+\phi_1)^2 -\frac{1}{L_x}\int_0^{L_x}dx\frac{\underline N_0 }{4\pi_{\gamma_0}}\tilde h M^2(\phi_0+\phi_1)^2
\ea
Coordinates in phase space are given by pairs of conjugate homogeneous and inhomogeneous fields, with Poisson brackets  
\bea \label{PB-red-old}
\{ \gamma_0,\pi_{\gamma_0}\}&=&\frac{1}{L_xL_yL_z}\, ,\quad \{ \tau_0,\pi_{\tau_0}\}=\frac{1}{L_xL_yL_z}\, ,\quad \{ \xi_0,\pi_{\xi_0}\}=\frac{1}{L_xL_yL_z}\, , 
\\ \nonumber \{ \phi_0,P_{\phi_0}\}&=&\frac{1}{L_xL_yL_z}\, ,\quad \{ \xi_1(x),\pi_{\xi_1}(x')\}=\frac{1}{L_yL_z}\delta(x-x')\, ,\quad \{ \phi_1(x),\pi_{\phi_1}(x')\}=\frac{1}{L_yL_z}\delta(x-x')\,.
\ea 
Besides, the dynamics is determined by the reduced total Hamiltonian  
\be\label{eq:red-totham-old}
\underline{\cal H}^{(2)} = N_0^x\underline V_x^{(2)}+\underline N_0\underline S^{(2)}.
\ee
with
\bea\nonumber
\underline S_0^{(2)} &=&L_xL_yL_z \bigg[\kappa\bigg(-2\pi_{\gamma_0}\pi_{\tau_0}\tau_0+\frac{\xi_0^2\pi_{\gamma_0}^2}{4\tau_0}+\pi_{\xi_0}^2\tau_0\bigg)+\frac{P^2_{\phi_0}}{2}\bigg]\\\label{eq:red-hamconst-old}
&&+L_yL_z \int_0^{L_x}dx\bigg[\kappa \bigg(\frac{\xi_1^2\pi_{\gamma_0}^2}{4\tau_0}+\pi_{\xi_1}^2\tau_0\bigg)+\frac{\tau_0(\xi_1')^2}{4\kappa}+\frac{P^2_{\phi_1}}{2}+\frac{\tau_0^2}{2}(\phi_1')^2+\tilde h V(\phi)\bigg],
\ea 
the reduced scalar constraint, and
\bea\label{eq:red-diff-old}
\underline V_x^{(2)} &=&L_yL_z \int_{0}^{L_x}\,dx\bigg[\pi_{\xi_1} \xi_1'+P_{\phi_1}\phi_1'\bigg],
\ea
the reduced diffeomorphism constraint. 

For convenience, we will introduce a canonical transformation in the homogeneous sector of the geometrical variables in order to cast it in a more familiar form in terms of the directional scale factors $a_1$, $a_2$ and $a_3$, and their conjugate momenta. Concretely,
\be\label{eq:hom-to-ai}
a_1^2 = e^{\gamma_0-\frac{\xi_0}{\sqrt{\tau_0}}-\frac{\xi^2_0}{4\tau_0}},\quad a_2^2 = \tau_0^2 e^{-\frac{\xi_0}{\sqrt{\tau_0}}}, \quad a_3^2 = e^{\frac{\xi_0}{\sqrt{\tau_0}}},
\ee
The conjugate momenta will then be given by
\bea\label{eq:hom-to-pi}
p_{a_1} &=& 2e^{-\tfrac{\gamma_0}{2}+\tfrac{\xi_0}{2\sqrt{\tau_0}}+\tfrac{\xi_0^2}{8\tau_0}} \pi_{\gamma_0},\\\nonumber
p_{a_2} &=& e^{\frac{\xi_0}{2\sqrt{\tau_0}}}\frac{\left(\xi_0\pi_{\xi_0}+2\tau_0\pi_{\tau_0}\right)}{2\tau_0},\\\nonumber
p_{a_3} &=&  2e^{-\frac{\xi_0}{2\sqrt{\tau_0}}}\left(\frac{\xi_0\pi_{\gamma_0}}{2\sqrt{\tau_0}}+\pi_{\gamma_0}+ \frac{\xi_0\pi_{\xi_0}}{4}+\pi_{\xi_0}\sqrt{\tau_0}+\frac{\tau_0\pi_{\tau_0}}{2}\right).
\ea
If we implement this canonical transformation, one can easily verify that the reduced spacetime line element is given by Eq. \eqref{eq:red-ds2}. Furthermore, the Poisson brackets of the homogeneous and inhomogeneous fields are those given in Eq. \eqref{PB-red}. Finally, the total Hamiltonian and the reduced constraints take the form given in Eqs. \eqref{eq:red-totham}, \eqref{eq:red-hamconst} and \eqref{eq:red-diff}, respectively.

\section{Isotropic sector}\label{app:iso-sys}

It is interesting to note that this cosmological setting admits isotropic and inhomogeneous solutions in some sense.\footnote{The stability of this sector with respect to small anisotropic deviations is an interesting question to be studied. One should expect that these solutions will not be stable, as in the homogeneous setting, if the scale factor is contracting. However, if the scale factor expanding, one should expect stability in most of the cases as long as the kinetic energy of the scalar field is not the dominant contribution in the Friedmann equation. In this case anisotropies dilute as fast as the matter content.} We propose to further reduce the system by taking the isotropic limit in the homogeneous sector of the model. Namely, we define the isotropic limit by means of the following identification of the three directional scale factors, i.e., $a_1=a_2=a_3$, and their momenta, as
\be\label{eq:iso-ai-a}
a_I = a,\quad p_{a_I}=\frac{p_a}{3}.
\ee
The Poisson brackets in the isotropic geometrical sector are\footnote{In terms of the original homogeneous variables, we have
\be\nonumber
\tau_0=a^2,\quad \gamma_0=\log a(4+\log a),\quad \xi_0=2a\log a,
\ee
and for the momenta
\be\nonumber
\pi_{\tau_0} = \frac{p_a}{6a}\left(2+\log a(1+\log a)\right),\quad
\pi_{\gamma_0} = \frac{ap_a}{6},\quad
\pi_{\xi_0} =  -\frac{p_a}{6}\left(1+\log a\right).
\ee}
\be
\{a,p_{a}\}=\frac{1}{L_xL_yL_z}.
\ee

Let us mention that the symmetry reduction \eqref{eq:iso-ai-a} leaves the symplectic structure of the inhomogeneous sector invariant. The reduced (inhomogeneous) spacetime metric of this isotropic subsector of the phase space is 
\be\label{eq:iso-gmunu}
ds^2=-\left[a^4 \underline N_0^2-(N_0^x+N_1^x)^2\right]\frac{\tilde h}{a^4}dt^2+2\frac{\tilde h}{a^4}(N^x_0+N_1^x)dxdt+\frac{\tilde h}{a^4}dx^2+a^2 \left[e^{-\frac{\xi_1}{a}}dy^2+e^{\frac{\xi_1}{a}}dz^2\right].
\ee
where $\tilde h=a^6e^{-\frac{1}{4a^2}\xi_1^2(x)-\frac{6}{ap_a}\left[P_{\phi_0}\phi_1(x)+\int_0^x \,d\tilde x\left(\pi_{\xi_1}(\tilde x) \xi_1'(\tilde x)+P_{\phi_1}(\tilde x)\phi_1'(\tilde x)\right)\right]}$ is the determinant of the spatial metric and with the shift vector determined by Eq. \eqref{eq:shift-red}, but restricted to the isotropic subsector, namely
\bea\label{eq:shift-iso}
[\underline N_1^x(x)]' = \frac{3\underline N_0}{2ap_a}\tilde h M^2(\phi_0+\phi_1)^2-\frac{1}{L_x}\int_0^{L_x}dx\frac{3\underline N_0}{2ap_a}\tilde h M^2(\phi_0+\phi_1)^2.
\ea
The total reduced Hamiltonian is then 
\be
{\cal H}^{(2)}_{\rm iso} = \underline N_0^x\underline V_{x}^{(2)}+\underline N_0S_{\rm iso}^{(2)}.
\ee
with the reduced scalar constraint
\begin{align}\nonumber
&S_{\rm iso}^{(2)} =L_xL_yL_z \bigg[-\frac{\kappa}{12}a^2p_a^2+\frac{P^2_{\phi_0}}{2}\bigg]+L_yL_z \int_0^{L_x}dx\,\bigg[\kappa \bigg(\frac{p_{a}^2}{144}\xi_1^2+a^2\pi_{\xi_1}^2\bigg)+\frac{a^2}{4\kappa}(\xi_1')^2\\\label{eq:iso-sca}
&+\frac{P^2_{\phi_1}}{2}+\frac{a^4}{2}(\phi_1')^2+\frac{1}{2}M^2\tilde h (\phi_0+\phi_1)^2\bigg],
\end{align}
The reduced diffeomorphism constraint remains the same as in Eq. \eqref{eq:red-diff}. As we see in Eq. \eqref{eq:iso-sca}, the presence of the potential breaks the linearity of the EOM of the inhomogeneous sector, as in the anisotropic case. However, we can still find a sector of the theory where linearity is (approximately) recovered. If the mass of the scalar field is not negligible, it is defined by  the regime in which the determinant of the metric $\tilde h$ contributes as a nearly homogeneous function. This implies
\be\label{eq:lin-reg-iso}
\chi_{\rm iso} =-\frac{1}{4a^2}\xi_1^2(x)-\frac{6}{ap_a}\left[P_{\phi_0}\phi_1(x)+\int_0^x \,d\tilde x\Big(\pi_{\xi_1}(\tilde x) \xi_1'(\tilde x)+P_{\phi_1}(\tilde x)\phi_1'(\tilde x)\Big)\right]\simeq 0,
\ee
for all $x$ and all times. This regime could correspond to spacetimes with a scale factor that has expanded enough and inhomogeneities have diluted sufficiently. 

Let us recall that the total reduced Hamiltonian was given by \eqref{eq:iso-sca}. However, if $\chi_{\rm iso}$ in Eq. \eqref{eq:lin-reg-iso} is negligibly small, the determinant of the spatial metric becomes a homogeneous function. Then, the reduced scalar constraint becomes quadratic in the inhomogeneities. Hence, we adopt a Fourier expansion. The reduced scalar constraint is now given by
\begin{align}\nonumber
&\tilde S_{\rm iso}^{(2)} =L_xL_yL_z \bigg[-\frac{\kappa}{12}a^2p_a^2+\frac{P^2_{\phi_0}}{2}+\frac{1}{2}a^6M^2\phi_0^2+ \sum_{n\neq 0}\,\bigg[\kappa \bigg(\frac{p_{a}^2}{144}|\tilde\xi_1|^2+a^2|\tilde\pi_{\xi_1}|^2\bigg)+\frac{a^2}{4\kappa}k_n^2|\tilde\xi_1|^2\\\label{eq:iso-sca-b}
&+\frac{|\tilde P_{\phi_1}|^2}{2}+\frac{a^4}{2}k_n^2|\tilde\phi_1|^2+\frac{1}{2}a^6M^2|\tilde\phi_1|^2\bigg].
\end{align}
The reduced diffeomorphism constraint remains the same as in Eq. \eqref{eq:red-diffb}. The total reduced Hamiltonian is then 
\be
{\tilde{\cal H}}^{(2)}_{\rm iso} = \underline N_0^x\underline V_{x}^{(2)}+\underline N_0\tilde{\underline S}_{\rm iso}^{(2)}.
\ee
Before we specify conditions for $\underline N_0^x$ and $\underline N_0$, we must recall that the inhomogeneous shift $N_1^x$ is not zero, although it can be very small if the universe dilutes inhomogeneities during expansion. Concretely, one can express Eq. \eqref{eq:shift-iso} as
\be
[\underline N_1^x(x)]' = \frac{3\underline N_0a^5}{2p_a}M^2\left(2\phi_0\phi_1+\phi_1^2-\frac{1}{L_x}\int_0^{L_x}dx\phi_1^2\right).
\ee
If we integrate it, 
\be\label{eq:shift-iso-final}
\underline N_1^x(x)-\underline N_1^x(0) = \frac{3\underline N_0a^5}{2p_a}M^2\int_{0}^xdx \left(2\phi_0\phi_1+\phi_1^2-\frac{1}{L_x}\int_0^{L_x}dx\phi_1^2\right),
\ee
Now, in order to specify $N_0^x$ and $\underline N_0$, we follow the construction presented in the main text. Namely, we adopt the gauge fixing condition $\Phi_{\tilde \phi_1}=\tilde \phi_1(k_1)-\alpha$ with $\alpha$ a constant complex number. Preservation of this gauge fixing condition amounts to 
\be
 N_0^x =\frac{i\underline N_0\tilde P_{\phi_1}(k_1)}{k_1 \alpha} .
\ee
The conjugate momentum $P_{\phi_1}(k_1)$ is obtained by solving the (global) diffeomorphism constraint as
\bea\label{eq:moment-cond-iso}
&&\alpha^* \tilde P_{\phi_1}(k_1) - \tilde P_{\phi_1}^*(k_{1})\alpha=\bigg[\tilde\pi^*_{\xi_1}(k_{1}) \tilde\xi_1(k_{1})-\tilde\pi_{\xi_1}(k_1) \tilde\xi_1^*(k_1)\bigg] \\\nonumber
&&+\sum_{n\neq0,1}n\bigg[\tilde\pi_{\xi_1}^*(k_n) \tilde\xi_1(k_n)-\tilde\pi_{\xi_1}(k_n) \tilde\xi^*_1(k_n)+\tilde P_{\phi_1}^*(k_n)\tilde\phi_1(k_n)-\tilde P_{\phi_1}(k_n)\tilde\phi^*_1(k_n)\bigg]. 
\ea
The right hand side is specified by the set of constants of motion
\be
O_{\tilde \xi_1}(k_n)=i\left(\tilde\pi_{\xi_1}^*(k_n) \tilde\xi_1(k_n)-\tilde\pi_{\xi_1}(k_n) \tilde\xi^*_1(k_n)\right),\quad O_{\tilde \phi_1}(k_n)=i\left(\tilde P_{\phi_1}^*(k_n) \tilde\phi_1(k_n)-\tilde P_{\phi_1}(k_n) \tilde\phi^*_1(k_n)\right),
\ee
for all $n$. We restrict $O_{\tilde \phi_1}(k_n)$ and $O_{\tilde \xi_1}(k_n)$ such that the right hand side of \eqref{eq:moment-cond-iso} equals zero. We get $(\alpha^* \tilde P_{\phi_1}(k_1) - \tilde P_{\phi_1}^*(k_{1})\alpha)=0$, which in general implies $\tilde P_{\phi_1}(k_1)=0$ (provided $\alpha$ is not real or purely imaginary) and therefore $ N_0^x =0$.

Therefore, for this gauge fixing condition, the scalar constraint \eqref{eq:iso-sca-b} rules the dynamics of the final reduced model, but with the initial data restricted by
\be
\sum_{n\neq 0}n\bigg[O_{\tilde \xi_1}(k_n)+O_{\tilde \phi_1}(k_n)\bigg]=0,\quad O_{\tilde \phi_1}(k_1) = 0,\quad \tilde \phi_1(k_1) = \alpha\neq 0.
\ee

Let us now set $\underline N_0 = a^{-3}$, which in the absence of inhomogeneities amounts to cosmic time. On the one hand, the equations of motion of the homogeneous sector are given in Appendix \ref{app:eom}. Besides, on solutions, the system must satisfy the constraint equation $\tilde S_{\rm iso}^{(2)}\sim 0$, which plays the role of the Friedmann equation of homogeneous scenarios. If we introduce the Hubble parameter $H=\frac{\dot a}{a}$, the energy density of the homogeneous part of the scalar field $\rho_{\phi_0} = \frac{P_{\phi_0}^2}{2a^6}+\frac{1}{2}M^2{\phi_0}^2$, and the following energy densities of the inhomogeneous sector
\bea\nonumber
&&\rho_{\phi_1} = \frac{1}{2 a^{2}}k_1^2|\alpha|^2+\frac{1}{2}M^2|\alpha|^2+\sum_{n\neq 0,1}\frac{|\tilde P_{\phi_1}(k_n)|^2}{2a^6}+\frac{1}{2}\frac{k_n^2}{a^2}|\tilde\phi_1(k_n)|^2+\frac{1}{2}M^2|\tilde\phi_1(k_n)|^2,\\\label{eq:rho_massive}
&& \rho_{\xi_1} = \sum_{k_n\neq 0}\frac{\kappa}{a^4} |\tilde\pi_{\xi_1}(k_n)|^2+\frac{1}{4\kappa}\frac{k_n^2}{a^4}|\tilde\xi_1(k_n)|^2,
\ea
one can see that
\be\label{eq:fried-inhom}
H^2=\frac{\kappa}{3}\left(1-\frac{1}{12a^2}\sum_{n\neq 0}|\tilde\xi_1(k_n)|^2\right)\left(\rho_{\phi_0}+\rho_{\phi_1}+\rho_{\xi_1}\right).
\ee
This equation indicates that the geometrical inhomogeneities must be such that 
\be\label{eq:collapse-cond}
\frac{1}{12a^2}\sum_{n\neq 0}|\tilde\xi_1(k_n)|^2\leq 1,
\ee
at all times. When the inequality saturates the Hubble parameter vanishes. 

On the other hand, the modes satisfy the second order differential equations
\bea\label{eq:xi1_iso}
&&\ddot{\tilde{\xi}}_1(k_n)+H\dot{\tilde{\xi}}_1(k_n)+\frac{k_n^2}{a^2}\tilde\xi_{1}(k_n)+\frac{\kappa^2 p_a^2}{36a^4}\tilde\xi_{1}(k_n)=0,\\\label{eq:phi1_iso}
&& \ddot{\tilde{\phi}}_1(k_n) +3H\dot{\tilde{\phi}}_1(k_n)+\frac{k_n^2}{a^2}\tilde{\phi_1}(k_n)+M^2\tilde{\phi_1}(k_n)=0,
\ea
with real coefficients. Now, given a complex solution $(\rm A)$ that can be split into tensor ${}^{(A)}u$ and scalar ${}^{(A)}v$ inhomogeneous modes, and given another solution $(B)$ (within the equivalence class of solutions that share the same scale factor $a(t)$ at all times), there are conserved inner products given by
\be\label{eq:ip-xi-iso}
\langle {}^{(A)}u(k_n,t),{}^{(B)}u(k_n,t)\rangle = i L_xL_yL_z\left[{}^{(A)}\tilde{\xi}_1^*(k_n,t){}^{(B)}\tilde{\pi}_{\xi_1}(k_n,t)-{}^{(A)}\tilde{\pi}^*_{\xi_1}(k_n,t){}^{(B)}\tilde{\xi}_1(k_n,t)\right],
\ee
for the tensor modes and 
\be\label{eq:ip-phi-iso}
\langle {}^{(A)}v(k_n,t),{}^{(A)}v(k_n,t)\rangle = i L_xL_yL_z\left[{}^{(A)}\tilde{\phi}_1^*(k_n,t){}^{(B)}\tilde{P}_{\phi_1}(k_n,t)-{}^{(A)}\tilde{P}^*_{\phi_1}(k_n,t){}^{(B)}\tilde{\phi}_1(k_n,t)\right],
\ee
for the scalar ones. We can then express the Fourier modes of both types of inhomogeneities as combinations of two complex solutions of positive norm $u_{\tilde \xi_1}(k_n,t)$ and $v_{\tilde \phi_1}(k_n,t)$ (and their complex conjugates) as
\bea\nonumber
&&\tilde \xi_1(k_n,t) = a_{\tilde \xi_1,u}(k_n)u_{\tilde \xi_1}(k_n,t)+a^*_{\tilde \xi_1,u}(-k_n)u^*_{\tilde \xi_1}(-k_n,t),\\\label{eq:Fmodes-iso}
&&\tilde \phi_1(k_n,t) = a_{\tilde \phi_1,v}(k_n)v_{\tilde \phi_1}(k_n,t)+a^*_{\tilde \phi_1,v}(-k_n)v^*_{\tilde \phi_1}(-k_n,t).
\ea

The complex constants of motion $a_{\tilde \xi_1,u}(k_n)$ and $a_{\tilde \phi_1,v}(k_n)$ determine the initial state of the Fourier modes. One can actually see that 
\be
O_{\tilde \xi_1}(k_n)=|a_{\tilde \xi_1,u}(k_n)|^2-|a_{\tilde \xi_1,u}(-k_n)|^2,\quad O_{\tilde \phi_1}(k_n)=|a_{\tilde \phi_1}(k_n,v)|^2-|a_{\tilde \phi_1,v}(-k_n)|^2.
\ee

Besides, two basis  of solutions $u_{\tilde \xi_1}(k_n,t)$ and $\tilde u_{\tilde \xi_1}(k_n,t)$ are related by the Bogoliubov transformations 
\be
\tilde u_{\tilde \xi_1} = \alpha_{\tilde u,u} \,u_{\tilde \xi_1}+\beta_{\tilde u,u} \, u^*_{\tilde \xi_1},
\ee
with 
\be
\alpha_{\tilde u,u} = \langle u_{\tilde \xi_1},\tilde u_{\tilde \xi_1}\rangle,\quad \beta_{\tilde u,u} = -\langle u^*_{\tilde \xi_1},\tilde u_{\tilde \xi_1}\rangle,
\ee
satisfying also $|\alpha_{\tilde u,u}|^2-|\beta_{\tilde u,u}|^2=1$. Finally,  the Bogoliubov transformation between two basis of solutions induces the transformation
\be
a_{\tilde \xi_1,u}(k_n) = \alpha \, a_{\tilde \xi_1,\tilde u}(k_n)+\beta^*  \,a^*_{\tilde \xi_1,\tilde u}(k_n).
\ee
This transformation leaves the modes on the left hand side of Eq. \eqref{eq:Fmodes-iso} (and their time derivatives) invariant. Consequently, it does not change the reaction of inhomogeneities on the homogeneous sector. This discussion for the inhomogeneous modes of $\xi_1$ also applies straightforwardly to the ones of the scalar field $\phi_1$.

\section{Hamilton's equations of motion}\label{app:eom}

Given the reduced scalar constraint \eqref{eq:red-totham-gfixed}, with ${\tilde{P}}_{\phi_1}(k_1)=0$ and ${\phi_1}(k_1)=\alpha$, cosmic time implies the choice of densitized lapse $\underline N_0=1/\sqrt{\tilde h}$, with $\tilde h=a^6$, where $a=(a_1a_2a_3)^{1/3}$ here represents the mean scale factor, we have 
\bea
&&\dot{a}_1 = \{a_1,\underline N_0\underline S^{(2)} \}|_{\underline N_0=a^{-3}}= \frac{\kappa a_1p_{a_1}}{2 a_2 a_3}-\frac{\kappa p_{a_2}}{2 a_3}-\frac{\kappa p_{a_3}}{2 a_2}+\frac{\kappa a_1p_{a_1}}{8 a_2^{2} a_3^{2}}\sum_{n\neq 0}|\tilde{\xi}_1|^2, \\
&&\dot{a}_2 = \{a_2,\underline N_0\underline S^{(2)} \}|_{\underline N_0=a^{-3}}= \frac{\kappa a_2 p_{a_2}}{2 a_1 a_3}-\frac{\kappa p_{a_1}}{2 a_3}-\frac{\kappa p_{a_3}}{2 a_1}, \\
&&\dot{a}_3 = \{a_3,\underline N_0\underline S^{(2)} \}|_{\underline N_0=a^{-3}}= \frac{\kappa a_3 p_{a_3}}{2 a_1 a_2}-\frac{\kappa p_{a_1}}{2 a_2}-\frac{\kappa p_{a_2}}{2 a_1},
\ea
\bea
\nonumber
&&\dot{{p}}_{a_1} = \{p_{a_1},\underline N_0\underline S^{(2)}\}|_{\underline N_0=a^{-3}} = \frac{\kappa p_{a_1} p_{a_2}}{2 a_1 a_3}+\frac{\kappa p_{a_1} p_{a_3}}{2 a_1 a_2}-\frac{\kappa p_{a_1}^{2}}{2 a_2 a_3}-M^{2} a_2 a_3 \phi^{2}\\
&&- \frac{\kappa p_{a_1}^{2}}{8 a_2^{2} a_3^{2}}\sum_{n\neq 0}|\tilde{\xi}_1|^2-M^{2} a_2 a_3 \sum_{n\neq 0}|\tilde{\phi}_1|^2, \\\nonumber
&&\dot{{p}}_{a_2} = \{p_{a_2},\underline N_0\underline S^{(2)}\}|_{\underline N_0=a^{-3}} = \frac{\kappa p_{a_1} p_{a_2}}{2 a_2 a_3}+\frac{\kappa p_{a_2} p_{a_3}}{2 a_1 a_2}-\frac{\kappa p_{a_2}^{2}}{2 a_1 a_3}-M^{2} a_1 a_3 \phi^{2} - \frac{\kappa }{a_1 a_2}\sum_{n\neq 0}|\tilde{\pi}_{\xi_1}|^2\\
&& -\frac{1}{4 \kappa a_1 a_2}\sum_{n\neq 0}k_n^2|\tilde{\xi}_1|^2+\frac{\kappa a_1p_{a_1}^{2}}{16 a_2^{3} a_3^{2}}\sum_{n\neq 0}|\tilde{\xi}_1|^2-\frac{a_3}{ a_1}\sum_{n\neq 0}k_n^2|\tilde{\phi}_1|^2-M^{2} a_1 a_3 \sum_{n\neq 0}|\tilde{\phi}_1|^2, \\\nonumber
&&\dot{{p}}_{a_3} = \{p_{a_3},\underline N_0\underline S^{(2)}\}|_{\underline N_0=a^{-3}} = \frac{\kappa p_{a_1} p_{a_3}}{2 a_2 a_3}+\frac{\kappa p_{a_2} p_{a_3}}{2 a_1 a_3}-\frac{\kappa p_{a_3}^{2}}{2 a_1 a_2}-M^{2} a_1 a_2 \phi^{2} - \frac{\kappa }{a_1 a_3}\sum_{n\neq 0}|\tilde{\pi}_{\xi_1}|^2\\
&& -\frac{1}{4 \kappa a_1 a_3}\sum_{n\neq 0}k_n^2|\tilde{\xi}_1|^2+\frac{\kappa a_1p_{a_1}^{2}}{16 a_2^{2} a_3^{3}}\sum_{n\neq 0}|\tilde{\xi}_1|^2-\frac{a_2}{a_1}\sum_{n\neq 0}k_n^2|\tilde{\phi}_1|^2-M^{2} a_1 a_2 \sum_{n\neq 0}|\tilde{\phi}_1|^2,
\ea
\bea
&&\dot{\phi}_0 = \{\phi_0,\underline N_0\underline S^{(2)}\}|_{\underline N_0=a^{-3}}=\frac{P_{\phi_0}}{a_1a_2a_3},\\ 
&&\dot{P}_{\phi_0} = \{{P}_{\phi_0},\underline N_0\underline S^{(2)}\}|_{\underline N_0=a^{-3}}=-a_1a_2a_3M^2\phi_0.
\ea
while for the inhomogeneities
\bea
&&\dot{\tilde{\xi}}_1(k_n) = \{\tilde{\xi}_1(k_n),\underline N_0\underline S^{(2)}\}|_{\underline N_0=a^{-3}}=\frac{2\kappa}{a_1}\tilde{\pi}_{\xi_1}(k_n),\\
&&\dot{\tilde{\pi}}_{\xi_1}(k_n) = \{{\tilde{\pi}}_{\xi_1}(k_n),\underline N_0\underline S^{(2)}\}|_{\underline N_0=a^{-3}}=-\frac{\kappa a_1p_{a_1}^2}{8a_2^2a_3^2}\tilde\xi_1(k_n)-\frac{k_n^2}{2\kappa a_1}\tilde\xi_1(k_n), \\
&&\dot{\tilde{\phi}}_1(k_n) = \{\tilde{\phi}_1(k_n),\underline N_0\underline S^{(2)}\}|_{\underline N_0=a^{-3}}=\frac{\tilde{P}_{\phi_1}(k_n) }{a_1a_2a_3},\\
&&\dot{\tilde{P}}_{\phi_1}(k_n) = \{{\tilde{P}}_{\phi_1}(k_n),\underline N_0\underline S^{(2)}\}|_{\underline N_0=a^{-3}}=-a_1a_2a_3M^2\tilde{\phi}_1(k_n)-\frac{a_2a_3}{a_1} k_n^2\tilde{\phi}_1(k_n).
\ea
On the other hand, in the limit in which the homogeneous sector is isotropic (see Appendix \ref{app:iso-sys}), we obtain
\bea
&&\dot{a} = \{a,\underline N_0\tilde S_{\rm iso}^{(2)}\}|_{\underline N_0=a^{-3}} = -\frac{\kappa p_a}{6 a}+\frac{\kappa p_a}{72 a^3}\sum_{n\neq 0}|\tilde \xi_1|^2, \\\nonumber
&&\dot{{p}}_{a} = \{p_a,\underline N_0\tilde S_{\rm iso}^{(2)}\}|_{\underline N_0=a^{-3}} = \frac{\kappa p_a^2}{6 a^2}-3M^2a^2{\phi}_0^2-\frac{2 \kappa}{a^2}\sum_{n\neq 0}|\tilde{\pi}_{\xi_1}|^2-\frac{1}{2\kappa a^2}\sum_{n\neq 0}k^2|\tilde \xi_1|^2\\
&&-2\sum_{n\neq 0}k^2|\tilde \phi_1|^2-3M^2a^2\sum_{n\neq 0}|\tilde \phi_1|^2, \\
&&\dot{\phi}_0 = \{\phi_0,\underline N_0\tilde S_{\rm iso}^{(2)}\}|_{\underline N_0=a^{-3}}=\frac{P_{\phi_0}}{a^3},\\ 
&&\dot{P}_{\phi_0} = \{{P}_{\phi_0},\underline N_0\tilde S_{\rm iso}^{(2)}\}|_{\underline N_0=a^{-3}}=-a^3M^2\phi_0.
\ea
while for the inhomogeneities
\bea
&&\dot{\tilde{\xi}}_1(k_n) = \{\tilde{\xi}_1(k_n),\underline N_0\tilde S_{\rm iso}^{(2)}\}|_{\underline N_0=a^{-3}}=\frac{2\kappa}{a}\tilde{\pi}_{\xi_1}(k_n),\\
&&\dot{\tilde{\pi}}_{\xi_1}(k_n) = \{{\tilde{\pi}}_{\xi_1}(k_n),\underline N_0\tilde S_{\rm iso}^{(2)}\}|_{\underline N_0=a^{-3}}=-\frac{\kappa p_a^2}{72a^3}\tilde\xi_1(k_n)-\frac{k^2}{2\kappa a}\tilde\xi_1(k_n), \\
&&\dot{\tilde{\phi}}_1(k_n) = \{\tilde{\phi}_1(k_n),\underline N_0\tilde S_{\rm iso}^{(2)}\}|_{\underline N_0=a^{-3}}=\frac{\tilde{P}_{\phi_1}(k_n) }{a^3},\\
&&\dot{\tilde{P}}_{\phi_1}(k_n) = \{{\tilde{P}}_{\phi_1}(k_n),\underline N_0\tilde S_{\rm iso}^{(2)}\}|_{\underline N_0=a^{-3}}=-a k^2\tilde{\phi}_1(k_n)-a^3M^2\tilde{\phi}_1(k_n).
\ea

\bibliography{ref}

\end{document}